\documentclass{article} % For LaTeX2e
\usepackage{iclr2027_conference,times}

\usepackage{amsmath,amsfonts,bm}

\def\eqref#1{equation~\ref{#1}}
\def\Algref#1{Algorithm~\ref{#1}}

\def\1{\bm{1}}

\DeclareMathAlphabet{\mathsfit}{\encodingdefault}{\sfdefault}{m}{sl}
\SetMathAlphabet{\mathsfit}{bold}{\encodingdefault}{\sfdefault}{bx}{n}

\usepackage{hyperref}
\usepackage{url}
\usepackage{booktabs}
\usepackage{hyperref}
\usepackage{url}
\usepackage{amsmath, amssymb, amsthm}
\usepackage{bm}
\usepackage{algorithm}
\usepackage{algorithm}
\usepackage{algpseudocode}
\usepackage{amsthm}
\usepackage{multirow}
\usepackage{algpseudocode}
\usepackage{tabularx}
\usepackage{graphicx}
\usepackage{array}
\usepackage{wrapfig}
\usepackage{pifont}
\usepackage[most]{tcolorbox}
\newtcblisting{promptbox}[1][]{
  listing only,
  breakable,
  enhanced,
  colback=gray!5,
  colframe=gray!45,
  boxrule=0.4pt,
  arc=2pt,
  left=6pt,
  right=6pt,
  top=5pt,
  bottom=5pt,
  listing options={
    basicstyle=\ttfamily\scriptsize,
    breaklines=true,
    columns=fullflexible,
    keepspaces=true,
    showstringspaces=false
  },
  title=#1,
  fonttitle=\bfseries\small,
  coltitle=black,
  attach boxed title to top left={yshift=-2mm,xshift=3mm},
  boxed title style={
    colback=white,
    colframe=gray!45,
    boxrule=0.4pt
  }
}
\newcolumntype{Y}{>{\centering\arraybackslash}X}

\providecommand{\1}{\mathbf{1}}

\providecommand{\Algref}[1]{Alg.~\ref{#1}}

\definecolor{draftcolor}{RGB}{135,55,25}

\title{HepUKE: Uncertainty-aware Knowledge Evolution for High Energy Physics Analysis Agents}

\author{Zijie Shang \\
Lanzhou University, China \\
Institute of High Energy Physics, Beijing, China \\
\texttt{shangzj18@lzu.edu.cn} \\
\And 
Ke Li\\
Institute of High Energy Physics, Beijing, China \\
\texttt{like@ihep.ac.cn} \\
\AND
Junkun Jiao\\
Jilin University, China\\
Institute of High Energy Physics, Beijing, China \\
\texttt{jiaojk1118@mails.jlu.edu.cn} \\
\And 
Shengchun Wang\\
Liaoning University, China\\
Institute of High Energy Physics, Beijing, China \\
\texttt{4032532138@smail.lnu.edu.cn}
\AND
Changzheng Yuan\\
Institute of High Energy Physics, Beijing, China \\
\texttt{yuancz@ihep.ac.cn}\\
\And
Xiongfei Wang\\
Lanzhou University, China \\
\texttt{wangxiongfei@lzu.edu.cn} \\
\AND
Beijiang Liu\\
Institute of High Energy Physics, Beijing, China \\
\texttt{liubj@ihep.ac.cn}\\
\And 
Weimin Song\\
Jilin University, China\\
\texttt{weiminsong@jlu.edu.cn} \\
\AND
Xiaoshen Kang\\
Liaoning University, China\\
\texttt{kangxiaoshen@lnu.edu.cn}
}
\iclrfinalcopy % Uncomment for camera-ready version, but NOT for submission.
\begin{document}
% \nolinenumbers

\maketitle

\begin{abstract}
Knowledge bases are essential to agents for high energy physics (HEP), supplying few-shot evidence for previously unseen tasks. Their principal utility lies in strategic guidance for novel analyses rather than ready-made answers or reusable code. Prevailing systems, however, treat all retrieved evidence as equally trustworthy and fail to accrue knowledge from the analyses under assistance. In this paper, inspired by the standard HEP treatment of uncertainty, we propose \textbf{\textsc{\underline{HepUKE}}}: an \textbf{\underline{U}}ncertainty-aware \textbf{\underline{K}}nowledge \textbf{\underline{E}}volution system for \textbf{\underline{HEP}} that repurposes information entropy to unify statistical uncertainty from LLM generation and systematic uncertainty from HEP methodological choices. Upon it, we construct a self-evolving domain-specific-language (DSL) knowledge base and release a large-scale expert-annotated HEP physics-analysis benchmark. Per-evidence confidence is quantified via the information entropy of sampled DSL generations: high-confidence precedents constrain program synthesis, whereas analyses lacking close precedents are validated and written back, sustaining continual knowledge evolution. To handle the highly redundant yet logically stringent nature of HEP knowledge, we introduce an early-stopping controller that fuses a backward stationarity test on the answer posterior with a forward marginal information gain estimate. On the 1,000-pair BESIII PhysicsQA benchmark, HepUKE reduces retrieval rounds by 43.0\% while preserving most answer quality, and on the out-of-domain HotpotQA benchmark, it reduces retrieval rounds by 14.5\% with a similar quality--efficiency trade-off. The code and datasets used in this work are available at \url{https://anonymous.4open.science/r/HepUKE-32FF}.

\end{abstract}

\section{Introduction}
\label{sec:intro}
High Energy Physics analyses require substantial domain knowledge, complex software, and large-scale data processing, making them promising targets for agent-based assistance. Recent work has explored HEP agents, retrieval-augmented generation (RAG), domain-specific languages (DSLs), coding agents, and multi-agent workflows~\citep{jiang2026agentichybridragevidencegrounded,atif2025celloaileveraginglargelanguage,suresh2024ragbasedsummarizationagentelectronion,jiao2026hepscriptdualusedslhumanai,desai2026rooagentllmagentrootbased,sun2026reliablellmgeneratedprogramshighenergy,cakir2026experimentallimitsphysicalinsight,moreno2026aiagentsautonomouslyperform,costa2026agentrivetautomatedproducingrivet,he2026drsaiagenticairealworld}. However, unseen analyses require more than static retrieval or one-shot generation: agents must extract transferable knowledge from validated precedents and assess its reliability. Existing HEP knowledge bases provide neither iterative retention of validated precedents nor confidence estimation for stored evidence, allowing uncalibrated procedures to introduce systematic bias during knowledge evolution.

Standard RAG implicitly treats semantic similarity as a proxy for evidential value~\citep{lewis2021retrievalaugmentedgenerationknowledgeintensivenlp,gao2024retrievalaugmentedgenerationlargelanguage}. This assumption is unreliable in HEP, where redundant documents may share terminology while differing in measurement type, physical relevance, or trustworthiness. Similar but distracting passages can degrade generation~\citep{deng2026computationgeometrymeaningidentity,Cuconasu_2024,shi2023largelanguagemodelseasily,yoran2024makingretrievalaugmentedlanguagemodels}, and external evidence may amplify model overconfidence~\citep{xiong2024llmsexpressuncertaintyempirical}. Methods such as Self-RAG and CRAG~\citep{asai2023selfraglearningretrievegenerate,yan2024correctiveretrievalaugmentedgeneration} assess generic textual credibility, but do not quantify HEP-specific physical relevance or generation-side confidence. Because an incorrect selection cut or fit configuration can propagate into the final physics result, reliability must be controlled on both sides of the RAG pipeline.

Information entropy provides a low-cost, training-free signal for LLM confidence and hallucination detection~\citep{kuhn2023semanticuncertaintylinguisticinvariances,quevedo2024detectinghallucinationslargelanguage}. Existing approaches mainly capture statistical uncertainty, namely output fluctuations under fixed evidence~\citep{manakul2023selfcheckgptzeroresourceblackboxhallucination}; systematic uncertainty from methodological choices and retrieved evidence remains largely unaddressed. We therefore use entropy as a two-channel gate: it filters evidence with low physical relevance or high systematic uncertainty and prevents uncertain generations from entering the knowledge base.

Inspired by the HEP treatment of uncertainty, we propose \textbf{\textsc{\underline{HepUKE}}}, an \textbf{\underline{U}}ncertainty-aware \textbf{\underline{K}}nowledge \textbf{\underline{E}}volution system for \textbf{\underline{HEP}}. HepUKE fuses statistical uncertainty in LLM generation with systematic uncertainty induced by HEP methodological choices into a confidence signal for reranking retrieved evidence. It further introduces (i) confidence-based reranking, (ii) confidence-gated, append-only deposition of verified precedents, and (iii) an early-stopping controller combining forward marginal information gain with backward stationarity detection. Together, these components support calibrated, self-evolving, and efficient RAG-based HEP analysis. Our contributions are:
\begin{itemize}
    \item We identify the limitations of similarity-only retrieval and propose HepUKE, which jointly controls evidence relevance, generation confidence, and knowledge evolution.
    \item We fuse statistical and systematic uncertainty into a unified confidence signal for DSL generation and confidence-gated precedent deposition.
    \item We develop an information-theoretic early-stopping controller that reduces redundant retrieval on HEP's highly repetitive corpus.
    \item Experiments on BESIII benchmarks show improvements in correctness and calibration with fewer retrieval rounds than standard RAG.
\end{itemize}

\section{Related Work}
\label{sec:related}

\textbf{HEP RAG and agents.} HEP agents and RAG systems have been applied to analysis assistance, DSL generation, and multi-agent workflows~\citep{jiang2026agentichybridragevidencegrounded,atif2025celloaileveraginglargelanguage,suresh2024ragbasedsummarizationagentelectronion,jiao2026hepscriptdualusedslhumanai,desai2026rooagentllmagentrootbased,sun2026reliablellmgeneratedprogramshighenergy,he2026drsaiagenticairealworld}. Most rely on static retrieval and lack explicit control of physical relevance, methodological uncertainty, and iterative retention of validated precedents. HepUKE addresses these limitations through confidence-gated retrieval and an append-only DSL knowledge base.

\textbf{Uncertainty estimation for LLMs.} Semantic entropy~\citep{kuhn2023semanticuncertaintylinguisticinvariances,quevedo2024detectinghallucinationslargelanguage} and multi-sample consistency methods such as SelfCheckGPT~\citep{manakul2023selfcheckgptzeroresourceblackboxhallucination} estimate uncertainty or hallucination risk, while external evidence can increase overconfidence~\citep{xiong2024llmsexpressuncertaintyempirical}. These methods primarily describe statistical variation under fixed evidence. HepUKE additionally models systematic uncertainty arising from methodological choices in retrieved HEP evidence and combines both components into answer confidence.

\textbf{Retrieval control, experience retention, and early stopping.} FLARE and Adaptive-RAG~\citep{jiang2023activeretrievalaugmentedgeneration,jeong2024adaptiveraglearningadaptretrievalaugmented} control retrieval using confidence or query complexity. Reflexion, Voyager, and ExpeL~\citep{shinn2023reflexionlanguageagentsverbal,wang2023voyageropenendedembodiedagent,zhao2024expelllmagentsexperiential} retain experience, but their admission rules rely mainly on task performance or self-evaluation rather than scientific validity. BOCPD~\citep{adams2007bayesianonlinechangepointdetection} and marginal information gain provide principled change and novelty signals. HepUKE combines backward BOCPD-based stationarity detection with forward JSD-based marginal information gain for externally computable early stopping, without reflection tokens or retraining.

\textbf{Gaps and positioning.} Existing methods do not jointly control evidence-side systematic uncertainty, generation-side statistical uncertainty, and scientifically validated knowledge evolution. HepUKE fills this gap with a unified confidence estimator, a dual gate spanning retrieval and precedent deposition, and an information-theoretic early-stopping controller.
\begin{figure}
    \centering
    \includegraphics[width=0.98\linewidth]{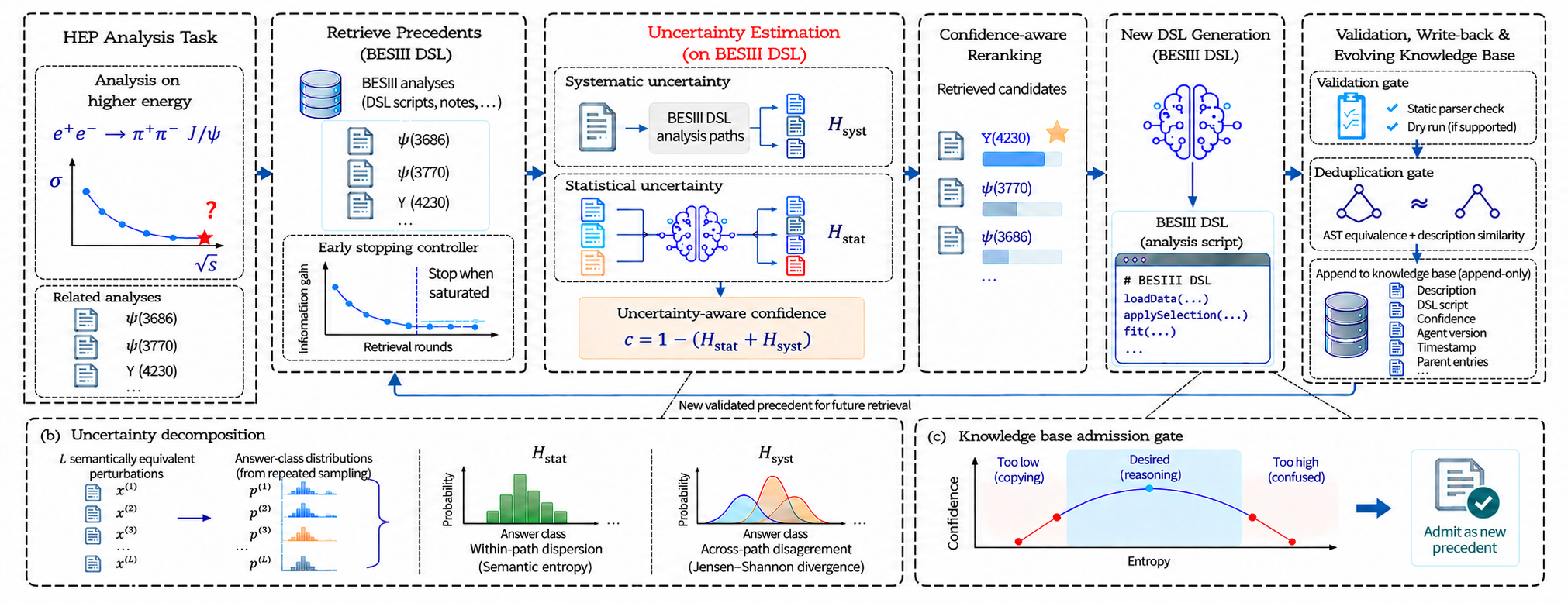}
\caption{\textbf{Overview of \textsc{HepUKE}.}
\textsc{HepUKE} performs uncertainty-aware precedent retrieval and BESIII DSL generation, using statistical and systematic uncertainty to guide confidence-based reranking and gated knowledge evolution. Bottom: (a) early stopping terminates saturated retrieval, (b) uncertainty is estimated across semantically equivalent DSL analysis paths, and (c) only validated precedents within the confidence gate are admitted to the evolving knowledge base.}    \label{fig:HEPUKE}
\end{figure}

\section{Preliminaries}
\label{sec:prelim}
\subsection{Problems}
Current HEP analysis agents face several challenges when reusing precedents across scientific documents and software workflows. First, retrieval relevance does not imply precedent usefulness: similar examples may introduce unsuitable topologies or numerical cuts. Second, systematic uncertainty introduced by methodology are not estimated, and sampling variation under a fixed instruction is easily confounded with sensitivity to controlled instruction changes. Third, program syntax is insufficient to establish scientific validity, necessitating explicit admission and reuse mechanisms. Fourth, a retrieval controller must allocate retrieval within a question, trading off a stability signal against the estimated value of another round. With redundant, evolving knowledge, over-retrieval wastes calls and hides correct information among near-duplicates. Existing RAG approaches, uncertainty quantification, and continuously evolving knowledge bases do not fully address these issues in scientific program synthesis.

\subsection{BESIII Experiment and \textsc{HepScript} Representation}
The Beijing Spectrometer III (BESIII)~\citep{Ablikim_2010} is an experiment at the large lepton collider Beijing Electron Positron Collider II (BEPCII)~\citep{Yu:IPAC2016-TUYA01} and has accumulated the world's largest $\tau$-charm dataset.
A complete analysis at BESIII typically spans detector simulation, event reconstruction, background suppression, event selection, statistical analysis, and systematic uncertainty evaluation, relying on large HEP frameworks and specific offline software.  This workflow is represented and driven by the \textsc{HepScritp}~\citep{jiao2026hepscriptdualusedslhumanai}, a DSL whose structured syntax enables parsing, comparison, and deduplication of analysis programs—properties unavailable for free-form scripts. Each DSL program encodes a set of observable analysis decisions as shown in~\Algref{alg:dsl}

\begin{algorithm}[t]
\caption{Generic HepScript template. }
\label{alg:dsl}
\begin{algorithmic}[1]
\State \textbf{Step 1 -- Datasets}
\State \quad \quad \quad $D_{\text{data}}\gets\textsc{DatasetManager.real\_data.find}(\text{``703\_4260''})$ \Comment{At 4.260 GeV}
\State \quad \quad \quad $D_{\text{incMC}}\gets\textsc{DatasetManager.inclusive\_mc.find}(\text{``703\_4260''})$
\State \textbf{Step 2 -- Decay card} $\mathcal{C}_{\text{sig}}$
\State \quad \quad \quad $\mathcal{C}_{\text{sig}}\gets\text{Decay}\ Y(4230)\!:\ \pi^{+}\pi^{-}J/\psi\ (\text{PHSP})$ \Comment{$e^{+}e^{-}\!\to\!\pi^{+}\pi^{-}J/\psi$}
\State \textbf{Step 3 -- Exclusive MC} $D_{\text{exMC}}$
\State \quad \quad \quad $D_{\text{exMC}}\gets\textsc{create\_exclusive\_mc}(\text{related}=D_{\text{data}},\ \text{events}=10^{5},\ \mathcal{C}_{\text{sig}})$
\State \textbf{Step 4 -- Algorithm config}
\State \quad\quad\quad $\mathcal{A}\gets\textsc{Analysis}(\text{name}=\text{``pipiJpsi''},\ \text{output}=\text{root})$
\State \textbf{Step 5 -- Selection Pipeline} \Comment{Analysis-specific}
\State \quad  \quad  \quad $\textsc{Track\ Selection}$
\State \quad  \quad  \quad $\textsc{Photon\ Selection}$
\State \quad  \quad  \quad $\textsc{Particle\ Identification}$
\State \quad  \quad  \quad $\textsc{Kinematic\ Fit}$
\State \textbf{Step 6 -- Note (un-expressible cut)}
\State \quad  \quad \quad $\textsc{Notes\ in\ context}$\Comment{free-text physics constraints}
\State \textbf{Step 7 -- Bind \& Execute}
\State \quad $\mathcal{A}.\textsc{bind}(D_{\text{data}},D_{\text{incMC}},D_{\text{exMC}},\mathcal{C}_{\text{sig}})$
\State \quad $\textsc{ROOT} \text{ Files} \gets \mathcal{A}.\textsc{execute\_on}([D_{\text{data}},D_{\text{incMC}}, D_{\text{exMC}}])$
\State \Return $\textsc{ROOT} \text{ Files}$
\end{algorithmic}
\end{algorithm}

\section{Methods}
\label{sec:method}
We present \textsc{HepUKE}, an uncertainty-aware knowledge evolution framework for HEP analysis.
It operates a retrieval loop on a shared knowledge base \(\mathcal{K} = \mathcal{K}_{\text{QA}} \cup \mathcal{K}_{\text{DSL}}\),
where \(\mathcal{K}_{\text{QA}}\) is a frozen retrieval index over publication chunks and
\(\mathcal{K}_{\text{DSL}} = \{e_i = (d_i, p_i, r_i)\}\) is a DSL library whose entries consist of a natural language description \(d_i\), a \textsc{HepScript} program \(p_i\) encoding observable analysis decisions, and its source chunk \(r_i\), thereby retaining provenance for later reuse and evolution.
The loop uses two complementary criteria:
answer confidence \(c\) governs \emph{\textbf{what is trustworthy}}, and retrieval saturation \(z\) governs \emph{\textbf{when to stop}}.
Our method consists of four components:
\begin{enumerate}
  \item an uncertainty-aware confidence estimator that factorizes $c$ into a statistical term $H_{\text{stat}}$ and a systematic term $H_{\text{syst}}$;
  \item a self-evolving DSL knowledge base, where \(c\) provides dual gating—weighting precedents and gating write-back—allowing \(\mathcal{K}_{\text{DSL}}\) to grow continuously from a static library;
  \item an early-stopping controller that infers $z$ from the dynamics of the posterior trajectory $\{p_t\}$ and determines the stopping round $\tau_t$;
  \item a retrieval loop that assembles the above modules into an auditable single session.
\end{enumerate}
\subsection{Uncertainty-Aware Confidence Estimator}
\label{sec:conf}

The estimator evaluates whether a candidate precedent reduces the uncertainty of the generated DSL relative to a matched baseline. It computes a statistical entropy term \(H_{\mathrm{stat}}\) from the language-model token probabilities and two complementary entropy summaries, \(H_{\mathrm{full}}\) and \(H_{\mathrm{sel}}\), over the full generated code and the \texttt{selection} module, respectively. Their formal definitions are given in Appendix~\ref{app:conf-method}. For query \(q\) and candidate \(k\), the systematic-uncertainty reduction relative to the matched baseline is defined as
\begin{equation}
\delta H_j(q,k) = H_j^{\mathrm{base}}(q) - H_j^{\mathrm{cand}}(q,k),
\qquad
j\in\{\mathrm{stat},\mathrm{full},\mathrm{sel}\}.
\label{eq:raw-entropy-reduction}
\end{equation}
A positive value indicates that the candidate reduces the corresponding uncertainty. The raw reductions are standardized globally over all query--candidate pairs, yielding \(\Delta H_{\mathrm{stat}}\), \(\Delta H_{\mathrm{full}}\), and \(\Delta H_{\mathrm{sel}}\). Although the statistical term is smaller than the other terms in the evaluated cells, it is retained in the score for methodological completeness.
The candidate score is defined as
\begin{equation}
\operatorname{score}(q,k) = \Delta H_{\mathrm{stat}}(q,k) + \alpha\,\Delta H_{\mathrm{full}}(q,k) + (1-\alpha)\,\Delta H_{\mathrm{sel}}(q,k),
\qquad
\alpha\in[0,1].
\label{eq:precedent-score}
\end{equation}
The candidate with the largest score is selected as the precedent. If \(k^\ast\) denotes the selected candidate, its confidence is
\begin{equation}
c(q) = \sigma\!\left( \operatorname{score}(q,k^\ast) \right).
\label{eq:confidence}
\end{equation}
Structural disagreement corresponds to the systematic term \(H_{\mathrm{syst}}\), nalogous to evaluating systematic uncertainties in HEP by varying methodological choices or analysis paths.
\(H_{\mathrm{syst,sel}}\) and \(H_{\mathrm{syst,glob}}\) are the JSDs of these symbols for the \texttt{selection} module and the full scripts, respectively, measuring structural disagreement across paths. Token-level predictive entropy reflects within-path statistical fluctuations, whereas structural disagreement reflects cross-path methodological sensitivity; the two signals are complementary.

\subsection{Self-Evolving DSL Knowledge Base}
\label{sec:kb}

The self-evolving DSL knowledge base extends each static entry
$e=(d_i,p_i,r_i)$ with a write-back confidence score $c_i$:
$e=(d_i,p_i,r_i,c_i)$. The original DSL script $r_i$ is retained, while provenance records the agent version, timestamp, and parent entries used for synthesis, allowing every entry to be traced to a human-written seed. Precedent retrieval uses the same hybrid dense--sparse index as the QA module and returns the five nearest candidates. A synthesized program is admitted only if it passes two gates. The validation gate requires successful static parsing and, when supported, a clean dry run. The deduplication gate rejects a candidate when an existing entry has an AST-equivalent program---under variable renaming, canonical ordering of commutable cuts and filters, and constant folding---and its
description similarity exceeds $\theta_{\text{dup}}$. This prevents
repeatedly writing the same program and skewing the retrieval distribution.
\subsection{Early-Stopping Controller}
\label{sec:ctrl}

The adaptive early-stopping controller terminates retrieval when the posterior has become sufficiently stable and another round is unlikely to provide useful information. After each round, a lightweight probe model constructs a posterior $p_t$ from the retrieval history. The controller combines a BOCPD-based stationarity score $\operatorname{Stab}_t$, a learned estimate $\hat g_\theta$ of the next-round information gain, the cumulative search cost, and an all-dry indicator:
\[
\begin{aligned}
\psi_t &=
\bigl[\hat g_\theta(\phi(\mathcal{H}_t)),
\operatorname{Stab}_t,t,C_t,\mathbb{1}_{\mathrm{dry}}\bigr],\\
z_t&=\sigma\!\left(\boldsymbol w^{\!\top}\psi_t+b\right),
\end{aligned}
\]
where $C_t$ is the cumulative number of search calls and
$\mathbb{1}_{\mathrm{dry}}$ indicates that the current round returned no documents. The gain predictor is trained by self-supervised regression on the Jensen--Shannon divergence between consecutive posteriors; further details are given in Appendix~\ref{app:ctrl-details}. Retrieval stops when $z_t>z_{\mathrm{th}}$ and $t\ge r_{\min}$, with $z_{\mathrm{th}}=0.7$ and $r_{\min}=2$. Stopping is blocked during cold-start rounds and whenever all subagents return no hits.

\section{Experiments}
In this section, we conduct comprehensive experiments to evaluate the three components in HepUKE -- the uncertainty-aware confidence estimator, the self-evolving knowledge base, and the saturation-gated stopping controller. Specifically, we aim to answer the following questions.\textbf{Q1} (Confidence Vs Relevance): How does uncertainty-aware confidence estimator perform compared to a cross-encoder relevance model? \textbf{Q2} (Self Evolution): How does confidence perform when used as the write-back gate of a self-evolving DSL knowledge base? \textbf{Q3} (Efficiency Analysis): How cost-efficient and time-efficient is the adaptive early-stopping controller in knowledge-redundant domains? \textbf{Q4} (Ablation Study): What contribution does each component of HepUKE make to the over all performance?

\subsection{Experimental Setting}
\paragraph{Datasets.} We employ three corpora, all pre-embedded in a single Milvus index that combines dense and sparse retrieval. The primary resource for synthesis and knowledge-evolution studies is BESIII DSL, a collection of 700 expert-authored analysis programs partitioned into a 500-program knowledge library \(\mathcal{K}_{\text{DSL}}\) and a held-out set of 200 targets; each entry is a triple $e=(d_i,p_i,r_i)$ linking a natural-language description, the executable DSL program, and its source publication chunk. For retrieval-loop evaluation we use BESIII‑PhysicsQA, which comprises 1000 manually reviewed question–answer pairs drawn from 700 BESIII publications, each declaring one of three scorers (exact match, numeric, or set-F1) and accompanied by gold evidence chunks in the frozen index. As an out-of-domain control for the stopping signal we include HotpotQA~\citep{yang2018hotpotqadatasetdiverseexplainable}, on which no confidence gating is applied. Paper isolation is enforced throughout, so a target’s own siblings are never returned during retrieval. We evaluate evidence selection on BESIII‑DSL, QA accuracy on BESIII‑PhysicsQA, and adaptive stopping on HotpotQA. The construction protocol of the BESIII-PhysicsQA benchmark is detailed in Appendix~\ref{app:benchmark}

\paragraph{Baselines. } We categorize all baselines into three groups according to the decision they replace. (i) \emph{Evidence selection}: We evaluate zero‑shot generation without retrieval (No‑RAG) as a lower anchor, the standard cross‑encoder RAG pipeline (Relevance top‑1), and the post‑hoc best candidate under gold as an upper anchor. (ii) \emph{Knowledge admission}: We compare a static library with no write‑back, a structural gate that enforces syntactic correctness and completeness, and a confidence threshold swept across a range. (iii) \emph{Stopping}: We contrast fixed‑round retrieval, native LLM self‑termination, and our saturation‑gated adaptive stopping. All configurations share an identical retriever, generator, and prompt, so that each comparison isolates the single decision under study.

\paragraph{Evaluation Metrics.} We evaluate our method using several metrics across two categories. For program synthesis and evidence selection, we adopt metrics defined as: (i) \emph{Strict match}, which checks whether datasets and decay cards are specified exactly; (ii) \emph{Structural match}, which measures operator-level and parameter-key consistency while deliberately ignoring concrete cut values; (iii) \emph{Phys-Acc}, which uses an LLM judge to assess whether the generated program encodes the same physics measurement as the ground truth, making this the only category that tolerates structural variation and is sensitive to numerical values.
For retrieval efficiency and knowledge-base health, we report: (i) \emph{Contain-Match Accuracy (Contain-Acc.)}, which checks if the correct answer appears in the generated response, (ii) \emph{Costs}, which include rounds, search calls, tokens, and time spent. For knowledge base evolution, the \emph{admission rate}, \emph{rejection breakdown}, and \emph{gate discrimination}---the downstream gap between admitted and rejected DSLs---to track the effect of write-back.

\paragraph{Implementations.} All methods share one retrieval stack: local \texttt{BGE-M3}
encoding (dense 1024-d and sparse) into a Milvus collections, hybrid recall, and
\texttt{bge-reranker-v2-m3} cross-encoder reranking, with $k=5$ throughout. All generation uses \texttt{deepseek-pro}: DSL synthesis at $T=0.7$, sampled $N=5$ per candidate. The saturation probe uses \texttt{deepseek-flash} at $\rm tempreture=0.7$. Experiments are conducted on the hardware configuration detailed in Appendix~\ref{sec:hardware}.
\begin{wrapfigure}[18]{r}{0.48\textwidth}
    \centering
    \setlength{\abovecaptionskip}{0pt}
    \includegraphics[width=\linewidth]{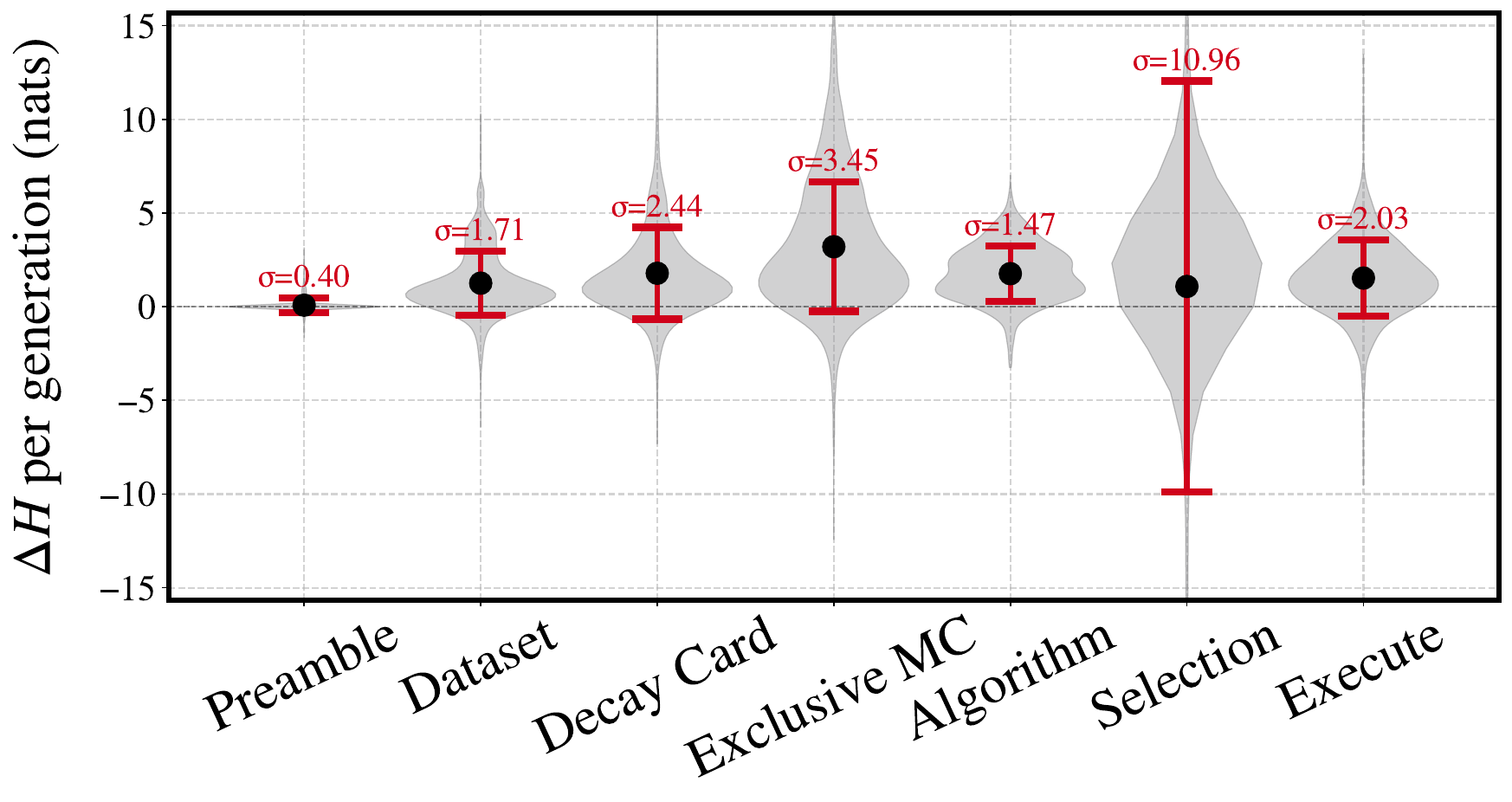}
    \caption{Distribution of the per-generation entropy reduction $\Delta H$ (in nats) across the seven DSL modules. Gray violins show the empirical distributions, black markers denote the means, and red bars denote one standard deviation, with the corresponding $\sigma$ values annotated above each module. Positive $\Delta H$ indicates reduced generation uncertainty relative to the matched baseline.}
    \label{fig:module}
\end{wrapfigure}
\subsection{Confidence vs Relevance (Q1)}
\label{sec:q1}
To address Q1, we fix the candidate pool, generator, and prompt, and vary only the strategy by which one of the \(k=5\) retrieved precedents is selected into the context. We evaluate no-RAG, standard RAG with a reranker, and different confidence-parameter configurations to assess the confidence estimator, and we further perform a module-wise uncertainty evaluation for HepScript. Our key findings are as follows:
% \begin{wrapfigure}[30]{r}{0.48\textwidth}
%     \centering
%     \includegraphics[
%     width=\linewidth,
%     trim=0 25 0 25,
%     clip]{figures/fig_dslgate_module_dH_weighted.pdf}
%     \caption{Distribution of the per-generation entropy reduction $\Delta H$ (in nats) across the seven DSL modules. Gray violins show the empirical distributions, black markers denote the means, and red bars denote one standard deviation, with the corresponding $\sigma$ values annotated above each module. Positive $\Delta H$ indicates reduced generation uncertainty relative to the matched baseline.}
%     \label{fig:module}
% \end{wrapfigure}
\paragraph{Obs.\ 1. Uncertainty-aware confidence  consistently outperforms relevance across protocols and metrics.} 
Direct DSL generation (No-RAG) performs weakest on the held-out BESIII-DSL targets, reaching only \(72.24\%\), \(89.60\%\), and \(58.54\%\) of the Oracle-normalized scores for strict match, structural match, and physical accuracy, respectively. Introducing a relevance-ranked precedent improves all three metrics to \(74.94\%\), \(89.72\%\), and \(62.46\%\). Confidence-aware selection yields further consistent gains, with the best configuration at \(\alpha=0.0\), where the three metrics reach \(76.08\%\), \(90.48\%\), and \(65.02\%\), all exceeding those of relevance Top-1. Notably, performance decreases monotonically as \(\alpha\) increases from \(0.0\) to \(1.0\), indicating that the selection-focused entropy reduction is more discriminative than the full-code entropy reduction; this is consistent with the observation that global structural uncertainty saturates easily. Nevertheless, although confidence-based selection substantially narrows the gap between model predictions and targets, it does not fully eliminate it.
\paragraph{Obs.\ 2. Uncertainty variation is concentrated in the task-specific selection segment.}
Decomposing \(\Delta H\) by DSL segment shows that \emph{Selection} has by far the widest distribution (\(\sigma=10.96\) nats) and substantial mass on both sides of zero, whereas the other segments are concentrated around positive reductions with \(\sigma=0.40\)--\(3.45\) nats. The reusable modules therefore benefit from precedents relatively consistently, with the largest mean reduction observed for \emph{Exclusive MC}. In contrast, the effect on the task-specific selection chain is strongly dependent on precedent compatibility and dominates the residual variation. Consistent with this localization, \(\Delta H_{\mathrm{sel}}\) alone achieves a \(39.8\%\) recovery rate, within \(1.4\) percentage points of the best fused setting.

\begin{table}[t]
\centering
\caption{Precedent selection on 200 BESIII-DSL samples. Scores normalized to Oracle (\%).}
\label{tab:gate}
\begin{tabular}{l c c c}
\toprule
Strategy & Strict Match & Struct Match& PhysicsAcc. \\
\midrule
\textsc{No-RAG}                             & 72.24 & 89.60 & 58.54 \\
\textsc{Relevance Top-1}                    & 74.94 & 89.72 & 62.46 \\
\textbf{Confidence} ($\alpha{=}0.0$)        & \textbf{76.08} & \textbf{90.48} & \textbf{65.02} \\
\textbf{Confidence} ($\alpha{=}0.3$)        & 75.59 & 89.88 & 64.18 \\
\textbf{Confidence} ($\alpha{=}0.5$)        & 75.17 & 89.41 & 64.42 \\
\textbf{Confidence} ($\alpha{=}0.7$)        & 74.76 & 88.83 & 64.27 \\
\textbf{Confidence} ($\alpha{=}1.0$)        & 74.57 & 88.77 & 63.79 \\
\textsc{Oracle}                             & 100.00 & 100.00 & 100.00 \\
\bottomrule
\end{tabular}
\end{table}

\subsection{Self Evolution (Q2)} 
\label{sec:q2}
To better understand whether confidence can serve as a dynamic library admission threshold, we conduct a direct evaluation of confidence as a thresholding criterion and, separately, examine whether the observed calibration structure can be converted into an improved selection rule.
\paragraph{Obs.\ 3. Uncertainty reduction has a sweet spot that only an interval gate can exploit.}
Table~\ref{tab:writeback} reports a sweep of $\tau_{\text{write}}$ over the write-back protocol. Under the one-sided gate, increasing $\tau_{\text{write}}$ from $0.5$ to $0.7$ reduces admissions from $191$ to $39$ and decreases admitted quality from $0.9301$ to $0.8861$, whereas rejected low-confidence programs remain high-quality ($0.8284$). The interval gate behaves differently: holding $\theta_{\text{write}}=0.5$ fixed and tightening $\theta_{\text{high}}$ from $0.95$ to $0.80$ raises admitted quality from $0.9352$ to $0.9457$, while rejecting $5$--$16$ high-confidence but low-quality programs ($0.4788$--$0.6435$)---precisely the transcription tail that a monotone gate would admit as its most confident candidates.

\begin{wrapfigure}[18]{r}{0.48\textwidth}
    \centering
    \setlength{\abovecaptionskip}{0pt}
    \includegraphics[width=\linewidth]{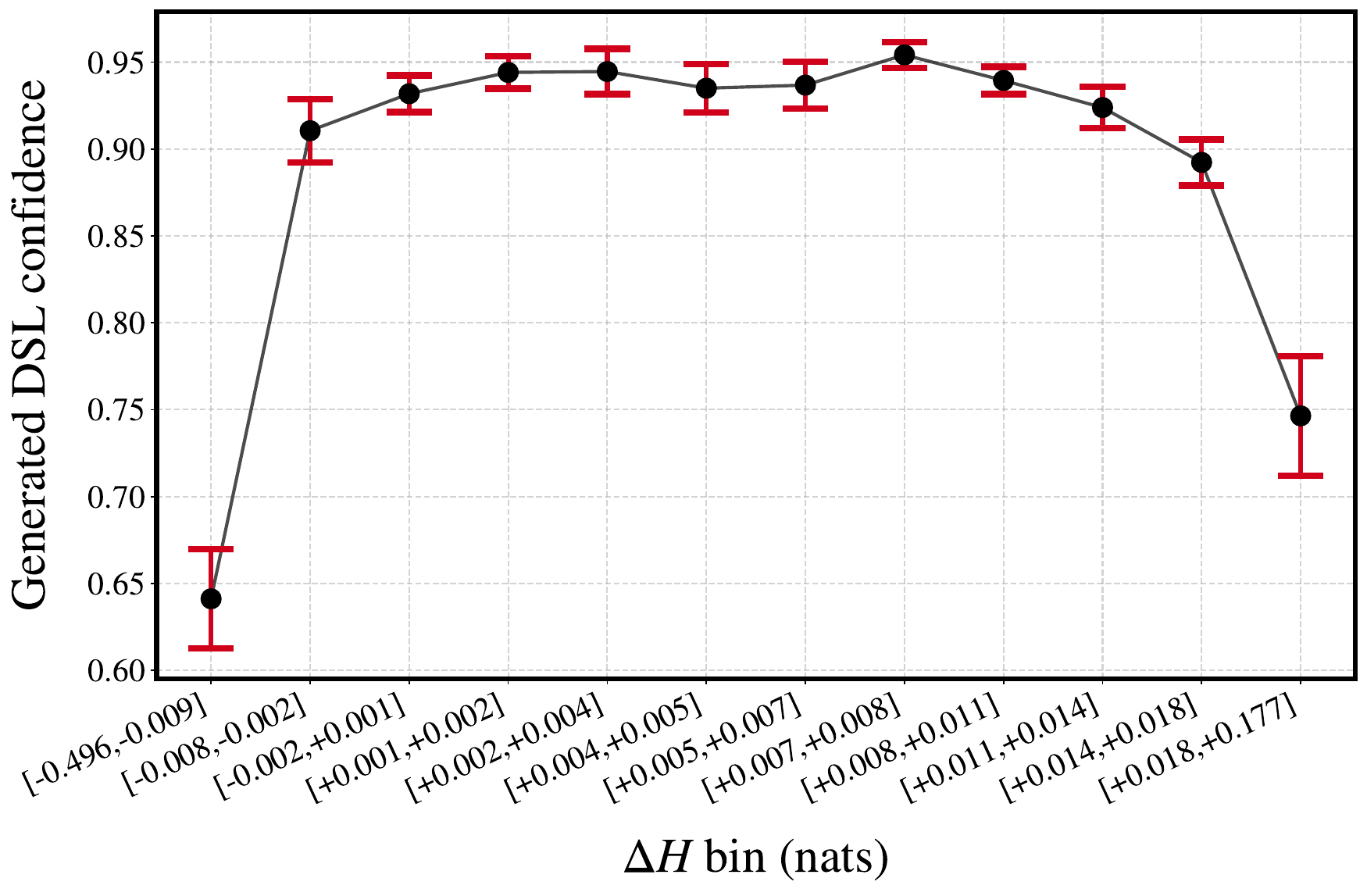}
    \caption{Binned relationship between per-generation entropy reduction $\Delta H_{\mathrm{bin}}$ and generated DSL confidence. Black markers show mean confidence, and red bars indicate within-bin uncertainty. Confidence is highest for moderate positive $\Delta H_{\mathrm{bin}}$ and decreases at both extremes.}
    \label{fig:calib}
\end{wrapfigure}

This yields an inverted-U profile, as shown in Figure~\ref{fig:calib}: quality peaks at $0.942$ for $\Delta H_{\text{sel}}\in[+0.005,+0.010)$ ($n=243$) and declines to $0.598$ for $\Delta H_{\text{sel}}\ge+0.04$ ($n=21$) and to $0.612$ for $\Delta H_{\text{sel}}\le-0.01$ ($n=79$). The two tails correspond to distinct failure modes: the right tail transcribes precedent-specific cut values, decay modes, and PID thresholds that do not transfer, whereas the left tail conflicts with the target and raises uncertainty. In the peak interval, the precedent fixes the program's structure while the model re-derives the target's specifics, rendering the reference informative without inducing copying. Accordingly, the knowledge base should admit a generated program only when its $\Delta H_{\text{sel}}$ falls within this sweet spot, thereby rejecting both tails---an interval threshold rather than a monotone confidence threshold.

\begin{table}[t]
\centering
\small
\caption{Write-back admission over 200 queries.}
\label{tab:writeback}
\begin{tabular}{@{}l r r r r r r r r r@{}}
\toprule
Gate & $\theta_{\mathrm{write}}$ & $\theta_{\mathrm{high}}$ &
\multicolumn{2}{c}{Admitted} & \multicolumn{2}{c}{Rej.\ low-conf} &
\multicolumn{2}{c}{Rej.\ high-conf} & Rej.\ struct \\
\cmidrule(lr){4-5}\cmidrule(lr){6-7}\cmidrule(lr){8-9}
 & & & $n$ & struct & $n$ & struct & $n$ & struct & struct \\
\midrule
Threshold & 0.5 & $\infty$ & 191 & 0.93 & 2   & 0.50 & 0   & $-$ & 0.52 \\
           & 0.6 & $\infty$ & 134 & 0.92 & 79  & 0.83 & 0   & $-$ & 0.51 \\
           & 0.7 & $\infty$ & 39  & 0.89 & 208 & 0.83 & 0   & $-$ & 0.52 \\
\midrule
Sweet Spot   & 0.5 & 0.95 & 189 & 0.94 & 2   & 0.50 & 5   & 0.48 & 0.5244 \\
           & 0.5 & 0.90 & 189 & 0.94 & 2   & 0.50 & 7   & 0.55 & 0.52 \\
           & 0.5 & 0.85 & 187 & 0.93 & 2   & 0.50 & 9   & 0.64 & 0.52 \\
           & 0.5 & 0.80 & 183 & \textbf{0.95} & 2   & 0.50 & 16  & 0.57 & 0.53 \\
\bottomrule
\end{tabular}
\end{table}

\subsection{Efficiency Analysis}
\label{sec:q3}
In order to evaluate the associated efficiency and cost implications of the early-stopping controller, we perform a dedicated analysis of the overhead and contain accuracy of HepUKE during the retrieval stage, comparing token usage and runtime with and without the early-stopping controller on the BESIII dataset. To further validate the domain transferability of the early-stopping method, we also evaluate retrieval overhead on the HotpotQA dataset. The results are presented in Table~\ref{tab:parity_cost}, and we summarize our key observations as follows:
\paragraph{Obs.\ 4. Early stopping exploits corpus redundancy for a favorable cost--accuracy trade-off.}
On BESIII PhysicsQA, \textsc{HepUKE} reduces retrieval rounds from 5.23 to 2.98 ($-43.0\%$), wall-clock time from 66.18 to 42.76\,s ($-35.4\%$), and total tokens from 69{,}001 to 38{,}041 ($-44.9\%$). Accuracy decreases by only 2.27 percentage points, from $50.31\%$ to $48.04\%$, retaining $95.5\%$ of native accuracy. In this highly redundant corpus, the controller therefore removes substantial repeated computation while preserving most answer quality. This trade-off is favorable when the application places greater weight on inference cost and latency than on recovering the final few percentage points of accuracy. The other RAG baselines have different retrieval objectives and are not directly comparable; notably, both Adaptive-RAG and FLARE obtain lower accuracies of $43.71\%$ and $25.35\%$, respectively. On HotpotQA, \textsc{HepUKE} reduces time by $18.2\%$ with only a 0.76-point decrease in answer F1, from $81.48\%$ to $80.72\%$.
  \begin{table}[t]
    \centering
    \small
    \setlength{\tabcolsep}{5pt}
    \caption{Answer quality and computational cost with and without the early-stopping controller on BESIII PhysicsQA and HotpotQA.}
    \label{tab:parity_cost}
    \begin{tabular}{l l c c c c c}
    \toprule
    & & & & \multicolumn{2}{c}{Tokens} & \\
    \cmidrule(lr){5-6}
    Benchmark & \makebox[24pt]{Method} & $N$ & Rounds & Prompt & Compl. & Accuracy (\%) \\
    \midrule
    \multirow{4}{*}{BESIII PhysicsQA}
      & Native       & \multirow{2}{*}{1,000}     & 5.23 & 66{,}017 & 2{,}984 & 50.31 \\
      & \textsc{HepUKE}  &                          &  2.98 & 36{,}158 & 1{,}883 & 48.13 \\
    % \cmidrule{2-8}
      & Adaptive-RAG & \multirow{2}{*}{1000}       &  2.00 & 14{,}206 &   929  & 43.71 \\
      & FLARE        &                            & 5.00 &  8{,}499 &   476  & 25.35 \\
    \cmidrule{1-7}
    \multirow{2}{*}{HotpotQA}
      & Native       & \multirow{2}{*}{1{,}000} &  2.14 & \multicolumn{2}{c}{---} & 81.48 \\
      & \textsc{HepUKE}  &                            &  1.83 & \multicolumn{2}{c}{---} & 80.72 \\
    \bottomrule
    \end{tabular}
    \vspace{4pt}

  \end{table}
 
\subsection{Ablation Study}
\label{sec:ablation}

Several component-level ablations have been presented above, including
uncertainty gating versus relevance top-1, early stopping versus a fixed
retrieval budget, and interval admission versus a monotone threshold. We
further examine the stability of confidence gating under retrieval-depth and
paraphrase perturbations.

\paragraph{Obs.\ 5. Confidence gating remains effective under retrieval and
prompt perturbations.}
Table~\ref{tab:perturb} compares No-RAG, relevance top-1, and confidence
gating under a deeper candidate pool ($k=10$) and paraphrased prompts. With
$k=10$, confidence gating provides consistent but moderate improvements over
relevance ranking in all three metrics, increasing Strict Match from
$75.7\%$ to $76.3\%$, Structural Match from $90.6\%$ to $90.8\%$, and
Phys-Acc from $62.6\%$ to $63.3\%$. Under paraphrase, the advantage is
larger: confidence gating reaches $76.0\%$, $90.7\%$, and $63.9\%$,
outperforming relevance top-1 by $0.8$, $0.4$, and $2.1$ percentage points
on Strict Match, Structural Match, and Phys-Acc, respectively. These results
indicate that the confidence signal remains useful when the retrieval pool is
expanded and when the prompt wording changes, with the largest benefit
appearing in physics-sensitive evaluation.

\begin{table}[t]
\centering
\small
\setlength{\tabcolsep}{4pt}
\caption{Results of perpetuated retrieval depth and paraphrase.
Each variant is evaluated under No-RAG (NR), relevance top-1 (Rel),
and confidence gating (Conf).}
\label{tab:perturb}
\begin{tabular}{l c c c c c c c c c}
\toprule
& \multicolumn{3}{c}{Strict (\%)}
& \multicolumn{3}{c}{Struct (\%)}
& \multicolumn{3}{c}{Phys-Acc (\%)} \\
\cmidrule(lr){2-4} \cmidrule(lr){5-7} \cmidrule(lr){8-10}
Perturbation & NR & Rel & Conf & NR & Rel & Conf & NR & Rel & Conf \\
\midrule
$k=3$
& 71.8 & \underline{76.2} & \textbf{76.7}
& 89.3 & \underline{91.0} & \textbf{91.0}
& 57.6 & \underline{62.4} & \textbf{63.4} \\
$k=10$
& 71.8 & \underline{75.7} & \textbf{76.3}
& 89.3 & \underline{90.6} & \textbf{90.8}
& 57.6 & \underline{62.6} & \textbf{63.3} \\
paraphrase
& 71.8 & \underline{75.2} & \textbf{76.0}
& 89.3 & \underline{90.3} & \textbf{90.7}
& 57.6 & \underline{61.8} & \textbf{63.9} \\
\midrule
\textsc{Oracle}
& \multicolumn{3}{c}{100}
& \multicolumn{3}{c}{100}
& \multicolumn{3}{c}{100} \\
\bottomrule
\end{tabular}
\end{table}

\section{Limitations}
\label{sec:limitations}

Our study has several limitations. First, entropy reduction is an operational proxy for precedent usefulness rather than a direct measure of physical correctness. The inverted-U pattern observed in the write-back experiment shows that very large reductions may reflect inappropriate copying of precedent-specific cuts or decay configurations. The interval gate mitigates this behavior, but its thresholds are calibrated only on the current BESIII-DSL setting.
Second, the DSL modules respond differently to precedent information: uncertainty reduction is concentrated in the task-specific \texttt{selection} module, while other modules exhibit smaller and more consistent changes. Our current estimator combines these signals globally and therefore does not fully exploit this module-specific structure. In addition, the experiments evaluate individual admission decisions rather than long-term knowledge evolution.

\section{Conclusion}
\paragraph{Summary.} HEP analysis agents typically start new pipelines with few-shot knowledge from dedicated bases, which offer strategic guidance rather than ready-made solutions. Existing bases, however, treat all retrieved items as equally trustworthy and learn nothing from the analyses they assist. We introduce \textsc{HepUKE}, which augments LLM statistical uncertainty with an HEP-specific systematic component from methodological choices. This estimator enables an append-only, self-evolving DSL library: high-confidence precedents constrain synthesis, while analyses without close precedents are validated and written back through an interval-shaped gate. For adaptive retrieval termination, we fuse a look-back stationarity test with a look-ahead marginal-information-gain estimate. On BESIII-DSL, confidence-aware selection yields further consistent gains, and confidence gating remains effective under retrieval and prompt perturbations. The interval gate improves admitted-program quality from $0.93$ to $0.95$ while rejecting the transcription tail. On the 1{,}000-item BESIII PhysicsQA benchmark, the controller cuts retrieval rounds by $43.0\%$ and prompt tokens by $44.9\%$ at essentially unchanged accuracy, and transfers to HotpotQA with cost savings preserved. The mechanism uses only standard LLM API outputs and is not tied to BESIII or a specific generator.
\paragraph{Future Work.} We will further explore how to incorporate physics physical plausibility and logical correctness into the framework, including executable DSL checks, numerical consistency tests, and explicit physics constraints. We will also extend the approach to additional HEP experiments and DSL representations.
\clearpage
\subsection*{AI Use Statement}

We used generative AI tools (large language model assistants) in this work. We disclose the following uses.

\textbf{Uses requiring disclosure:} synthetic data generation; implementation of methods (code writing); dataset cleaning; and manuscript polishing. Details of the AI models, prompts, generation and cleaning pipelines, and human verification used for synthetic data generation and dataset cleaning are provided in Appendix~\ref{app:phys-acc-prompt} and~\ref{app:benchmark}.

\textbf{Uses recommended for disclosure:} language polishing; figure preparation; plotting and evaluation code; literature search; and brainstorming. Some categories overlap --- for example, language polishing falls under manuscript polishing, and plotting/evaluation code falls under code writing.

Generative AI tools were not use for theoretical model construction or hypothesis generation. We also did not use AI for proving mathematical claims; this study is empirical and contains no formal theorems or proofs. The mathematical tools used in this work (information entropy, posterior stationarity tests, marginal information gain) are established methodological instruments, not claims requiring proof assistance.

All AI-assisted work was reviewed by the authors. The system codebase was written and refactored with the assistance of an LLM coding assistant. All such code was reviewed by the authors and covered by a unit-test suite; several LLM-introduced defects were identified and corrected through testing and manual review. We manually verified the existence and accuracy of each reference and did not retain any LLM-provided citation that we had not personally verified.

The PDF and supplementary materials contain no hidden instructions or invisible text directed at automated review systems. We assume responsibility for the final content of this work, including text, claims, and artifacts produced with the aid of generative AI.

\section*{Reproducibility Statement}
\label{sec:reproducibility}

We provide an anonymous code repository containing code, configurations, prompts, and evaluation scripts at \url{https://anonymous.4open.science/r/HepUKE-32FF}. Details of the uncertainty measures and controller are fully specified in Appendices~\ref{app:conf-method} and~\ref{app:ctrl-details}. BESIII-PhysicsQA is constructed following the deterministic seven-step protocol in Appendix~\ref{app:benchmark}. The physics-accuracy evaluation follows Appendix~\ref{app:phys-acc}. Hardware and software configurations are reported in Appendix~\ref{sec:hardware}.


\bibliography{iclr2027_conference}
\bibliographystyle{iclr2027_conference}
\clearpage
\appendix
\appendix
\section{Operational Uncertainty Features}
\label{app:conf-method}

\subsection{Scope and Notation}

The uncertainty measures used in this work are operational proxies for predictive uncertainty and structural disagreement. They are computed from repeated DSL generations under a baseline condition or a candidate-reference condition. They do not construct an answer-class mixture and do not estimate a signed systematic bias relative to a ground-truth analysis.

Let \(r\) denote a generation condition, such as the baseline condition or the condition in which candidate precedent \(k\) is provided. For each condition, multiple DSL samples are generated independently. The resulting quantities are first computed within each cell and are then compared between matched baseline and candidate cells.

\subsection{Token-Level Predictive Entropy}

For a generated token \(t\), let \(p_{t,i}\) denote the probability of the \(i\)-th token among the reported top-20 log-probability entries. The probability mass outside the reported entries is
\begin{equation}
m_t = 1-\sum_{i=1}^{20}p_{t,i}.
\end{equation}
The token-level predictive entropy is defined by
\begin{equation}
H_{\mathrm{tail}}(t) = -\sum_{i=1}^{20}p_{t,i}\ln p_{t,i} - m_t\ln m_t,
\label{eq:tail-entropy}
\end{equation}
with the convention \(0\ln 0=0\). The residual term accounts for the probability mass not included in the top-20 entries.
Let \(\mathcal{T}_{\mathrm{code}}^{(n,r)}\) be the set of kept code tokens in sample \(n\) under condition \(r\), excluding tokens assigned to the \texttt{note} module. The code-only entropy of this sample is
\begin{equation}
H_{\mathrm{stat,code}}^{(n,r)} = \frac{1}{\left|\mathcal{T}_{\mathrm{code}}^{(n,r)}\right|} \sum_{t\in\mathcal{T}_{\mathrm{code}}^{(n,r)}} H_{\mathrm{tail}}(t).
\label{eq:h-stat-code-sample}
\end{equation}
For a cell containing \(N\) generated samples, the corresponding cell-level quantity is
\begin{equation}
H_{\mathrm{stat,code}}^{(r)} = \frac{1}{N} \sum_{n=1}^{N} H_{\mathrm{stat,code}}^{(n,r)}.
\label{eq:h-stat-code-cell}
\end{equation}
For candidate comparison, two additional token-entropy summaries are computed. Let \(\mathcal{T}_{\mathrm{full}}^{(n,r)}\) denote all kept code tokens, including tokens in \texttt{note} modules, and let \(\mathcal{T}_{\mathrm{sel}}^{(n,r)}\) denote the subset belonging to the \texttt{selection} module. Then
\begin{align}
H_{\mathrm{full}}^{(n,r)} &= \frac{1}{\left|\mathcal{T}_{\mathrm{full}}^{(n,r)}\right|} \sum_{t\in\mathcal{T}_{\mathrm{full}}^{(n,r)}} H_{\mathrm{tail}}(t), \label{eq:h-full-sample} \\
H_{\mathrm{sel}}^{(n,r)} &= \frac{1}{\left|\mathcal{T}_{\mathrm{sel}}^{(n,r)}\right|} \sum_{t\in\mathcal{T}_{\mathrm{sel}}^{(n,r)}} H_{\mathrm{tail}}(t). \label{eq:h-sel-sample}
\end{align}
Their cell-level means are
\begin{equation}
H_j^{(r)} = \frac{1}{N} \sum_{n=1}^{N} H_j^{(n,r)}, \qquad j\in\{\mathrm{full},\mathrm{sel}\}.
\label{eq:cell-entropy-mean}
\end{equation}

\subsection{Structural Disagreement from Canonicalized DSL}

Structural disagreement is measured from discrete representations of the generated DSL rather than from token probability distributions. Complete \texttt{note} blocks are first removed from each generated program. The remaining source is canonicalized by removing comments, replacing string and numerical literals with fixed placeholders, collapsing whitespace, and discarding empty lines. The canonicalized source is then mapped to a discrete structural symbol by a deterministic hash function.

Let \(s_{\mathrm{sel}}^{(n,r)}\) denote the symbol obtained from the canonicalized \texttt{selection} module in sample \(n\), and let \(s_{\mathrm{glob}}^{(n,r)}\) denote the symbol obtained from the complete note-stripped program. For either scope \(a\in\{\mathrm{sel},\mathrm{glob}\}\), define the empirical symbol distribution
\begin{equation}
\widehat p_a^{(r)}(v) = \frac{1}{N} \sum_{n=1}^{N} \mathbf{1}\!\left[ s_a^{(n,r)}=v \right].
\label{eq:empirical-symbol-distribution}
\end{equation}

The structural disagreement quantity is
\begin{equation}
H_{\mathrm{syst},a}^{(r)} = -\sum_v \widehat p_a^{(r)}(v) \ln \widehat p_a^{(r)}(v), \qquad a\in\{\mathrm{sel},\mathrm{glob}\}.
\label{eq:structural-entropy}
\end{equation}

The implementation evaluates this quantity through a generalized Jensen--Shannon divergence routine. Because each sample contributes one discrete symbol, its within-sample distribution is one-hot and has zero entropy. Consequently, the implemented quantity reduces exactly to the empirical entropy in Eq.~\eqref{eq:structural-entropy}. It therefore measures disagreement among canonicalized DSL structures, rather than disagreement among token probability distributions.

\subsection{Baseline--Candidate Deltas and Confidence Score}

For a query identifier \(q\), let \(X^{\mathrm{base}}(q)\) be a cell-level feature computed under the baseline condition and \(X^{\mathrm{cand}}(q,k)\) the corresponding feature computed when candidate \(k\) is used as the reference. The raw candidate-induced change is defined by
\begin{equation}
\delta X(q,k) = X^{\mathrm{base}}(q) - X^{\mathrm{cand}}(q,k).
\label{eq:raw-delta}
\end{equation}
A positive value indicates that the candidate reduces the corresponding entropy or structural-disagreement measure.

For each feature \(X\), the raw deltas are standardized globally over all query--candidate pairs:
\begin{equation}
\Delta X(q,k) = \frac{\delta X(q,k)-\mu_X}{\sigma_X},
\label{eq:global-zscore}
\end{equation}
where \(\mu_X\) and \(\sigma_X\) are the mean and standard deviation of the complete set of raw deltas for feature \(X\). No task-specific or dataset-specific grouping is applied.

The candidate score combines the standardized full-code and selection-module entropy reductions:
\begin{equation}
\operatorname{score}(q,k) = \Delta H_{\mathrm{stat}}(q,k) + \alpha\,\Delta H_{\mathrm{full}}(q,k) + (1-\alpha)\,\Delta H_{\mathrm{sel}}(q,k),
\qquad
\alpha\in[0,1].
\label{eq:precedent-score}
\end{equation}
The selected candidate is
\begin{equation}
k^\ast = \operatorname*{arg\,max}_{k}\operatorname{score}(q,k),
\end{equation}
and its confidence is
\begin{equation}
c(q) = \sigma\!\left(\operatorname{score}(q,k^\ast) \right).
\label{eq:confidence}
\end{equation}
Thus, the confidence score increases when the selected precedent produces a larger standardized reduction in full-code and selection-module predictive entropy relative to the matched baseline.

\subsection{Controller Implementation Details}
\label{app:ctrl-details}

The probe posterior comes from a single short forward pass over $\mathcal{H}_t$ and a softmax over the top-$B$ answer-token log-probabilities ($B=20$). Its normalized certainty is
\[
u_t=1-\frac{H(p_t)}{\log B}.
\]
BOCPD uses a one-dimensional Beta observation model with hazard $H=0.1$, prior $(\alpha_0,\beta_0)=(1,1)$, and run-length pruning threshold $10^{-6}$. The run-length posterior is updated in the standard way, with sufficient statistics $\alpha_r\leftarrow\alpha_r+u_t$ and $\beta_r\leftarrow\beta_r+1-u_t$; a new run restarts from the prior. Observations and Beta-density endpoints are clipped to $[0,1]$ and $10^{-12}$, respectively. If normalization underflows, the posterior resets to the prior and $\operatorname{Stab}_t=0$.

The predictor input is a fixed feature vector combining round index, posterior statistics, document-set novelty, and recent realized-gain summaries:
\[
\phi(\mathcal{H}_t)=(t,H(p_t),\max p_t,\operatorname{Var}(p_t),
n_{\mathrm{unique}},1-J_{\mathrm{doc}},
\overline g_{3},\operatorname{Var}(g_{3}),
\operatorname{slope}(g_{3}),
1-\min(1,n_{\mathrm{unique}}/5)),
\]
where $J_{\mathrm{doc}}$ is the Jaccard similarity between current and historical document sets. The self-supervised target is the JSD between consecutive posteriors,
\[
g_{t+1}=\operatorname{JSD}(p_{t+1}\|p_t),
\]
using the shorter prefix when posterior lengths differ.

The fusion vector is
\[
\psi_t=(\hat g_\theta(\phi(\mathcal{H}_t)),
\operatorname{Stab}_t,t,C_t,\mathbb{1}_{\mathrm{dry}}),
\]
where $C_t$ is the cumulative number of search-tool calls and $\mathbb{1}_{\mathrm{dry}}$ is the all-dry indicator. The latter is also a hard safety condition, while its learned coefficient is zero. We search a single-stage grid of $14{,}000$ candidates over weights, bias, and threshold, subject to $R_{\mathrm{gold}}\ge R_{\mathrm{gold}}^{\mathrm{base}}-0.02$, and rank them lexicographically by rounds saved and then $R_{\mathrm{gold}}$, where
\[
R_{\mathrm{gold}}=
\mathbb{1}\!\left[
\mathrm{gold\ titles}\subseteq
\bigcup_{t\le t_{\mathrm{stop}}}\mathrm{doc\_ids}_t
\right]
\]
is a lower bound on answer accuracy, since retrieving the gold documents is necessary but not sufficient for an exact answer.

\section{Experimental Hardware Configuration}
\label{sec:hardware}
\begin{table}[H]
\centering
\small
\caption{Hardware and software configuration used in the experiments.}
\label{tab:hardware}
\begin{tabular}{ll}
\toprule
\textbf{Component} & \textbf{Configuration} \\
\midrule
Architecture & x86\_64 \\
CPU & 2 $\times$ Intel Xeon Gold 6430 \\
CPU cores & 64 physical cores (32 per socket) \\
Maximum CPU frequency & 3.4\,GHz \\
GPU & NVIDIA GeForce RTX 5090 \\
GPU memory & 32\,GB \\
NVIDIA driver & 580.95.05 \\
CUDA & 13.0 \\
L3 cache & 120\,MB \\
NUMA nodes & 2 \\
\bottomrule
\end{tabular}
\end{table}

\section{Physics-Acc. Evaluation Protocol}
\label{app:phys-acc}

We evaluate the generated HepScripts using a strict,
dimension-wise comparison between a reference specification and a candidate specification. The evaluation is performed by an LLM prompted to act as a senior BESIII analyst. It compares the physics content of the two DSL programs while ignoring variable names, comments, indentation, statement ordering, and free-text notes.
The comparison covers seven dimensions: decay chain and final state, dataset and energy, topology and tagging, track and photon selection, particle identification, fits and constraints, and MC configuration and decay card. For each dimension, the evaluator records the evidence, assigns a score, and lists substantive and minor differences. A dimension is assigned \texttt{null} only when the reference does not address that aspect. If the candidate omits a
requirement present in the reference, it receives a score of zero for the corresponding dimension.

\begin{table}[t]
\centering
\small
\caption{Dimensions used for physics-Acc. evaluation.}
\label{tab:physacc-dimensions}
\begin{tabular}{l p{0.68\linewidth}}
\toprule
\textbf{Dimension} & \textbf{Compared content} \\
\midrule
Decay chain and final state
& Parent particle, complete decay chain, and reconstructed final-state particles. \\
Dataset and energy
& Real-data and inclusive-MC samples, dataset identifiers, and centre-of-mass energy. \\
Topology and tagging
& Single- or double-tag topology, tagged and signal sides, and recoil-mass or direct reconstruction. \\
Track and photon selection
& Track quality, multiplicity, net charge, photon thresholds, timing, isolation, and photon multiplicity. \\
PID
& Identification method, likelihood or probability cuts, tested hypotheses, and particle multiplicities. \\
Fits and constraints
& Fit type, fitted particle list, constraints, and $\chi^2$ requirements. \\
MC and decay card
& Exclusive-MC setup, EvtGen decay chain, generator models, and requested event count. \\
\bottomrule
\end{tabular}
\end{table}
\begin{table}[t]
\centering
\small
\caption{Scoring criteria for physics-Acc. evaluation.}
\label{tab:physacc-criteria}
\begin{tabular}{c p{0.72\linewidth}}
\toprule
\textbf{Score} & \textbf{Interpretation} \\
\midrule
100 & Substantively identical; differences are purely notational. \\
80 & Physically equivalent; only minor threshold differences are present,
     with all relative deviations below 20\%. \\
60 & Same overall intent, but one substantive difference or several minor
     differences that may visibly change the selected sample. \\
40 & Partially overlapping; the dimension is addressed but contains multiple
     substantive errors. \\
20 & Addressed only nominally; most of the specified content is incorrect. \\
0 & The dimension is absent from the candidate or directly contradicts the
    reference. \\
\bottomrule
\end{tabular}
\end{table}

Any difference in a discrete physics choice is treated as substantive. This
includes the parent or daughter particle, decay mode, tagged side, PID
hypothesis set, fit type, generator model, energy point, or dataset identifier.
For continuous requirements, a relative deviation greater than 20\% is
substantive, whereas a deviation below 20\% is minor. An omitted requirement
is always treated as substantive.

The \emph{Phys-Acc} score is assigned by an LLM judge according to the above
protocol. It measures whether the candidate specification encodes the same
physics measurement as the reference, while allowing physically equivalent
differences in syntax and structure. The judge returns a JSON object containing
the summaries, identified differences, dimension-wise scores, a fatal-mismatch
flag, and a final verdict. A fatal mismatch is reported when the candidate
describes a fundamentally different parent particle or decay chain that cannot
be reconciled by adjusting numerical thresholds.

\subsection{LLM Judge Prompt}
\label{app:phys-acc-prompt}

The following prompt is used to evaluate the \emph{Phys-Acc} of a candidate HepScript.

\begin{promptbox}[\emph{Phys-Acc} Evaluation Prompt]
You are a senior BESIII collaboration analyst performing a strict technical
review. You are given two BOSS analysis specifications written in a Ruby
domain-specific language: the REFERENCE, which is the accepted specification,
and the CANDIDATE, which is to be reviewed.

Be strict. Do not assume that a difference is acceptable. A candidate that
selects different events is a failure even if it appears superficially similar.

Any difference in a discrete choice is substantive, including the particle,
decay mode, tagged side, PID hypothesis set, fit type, generator model, energy
point, or dataset identifier. For continuous thresholds, a relative deviation
above 20\% is substantive and a deviation below 20\% is minor. An omitted
requirement imposed by the reference is substantive.

Ignore variable names, comments, indentation, statement ordering, and the
content of free-text note blocks.

Evaluate the following dimensions: decay chain and final state; dataset and
energy; topology and tagging; track and photon selection; PID; fits and
constraints; and MC configuration and decay card. Use the scoring criteria
defined in Table~\ref{tab:physacc-criteria}. Return null only when a dimension
is not addressed by the REFERENCE.

Respond with a single JSON object and nothing else. Fill the fields in order:
evidence, per-dimension scores, and summary fields.

{
  "reference_summary": "<1-2 sentences>",
  "candidate_summary": "<1-2 sentences>",
  "substantive_differences": ["<one difference per item>"],
  "minor_differences": ["<one difference per item>"],
  "scores": {
    "decay_chain_and_final_state": <0-100 or null>,
    "dataset_and_energy": <0-100 or null>,
    "topology_and_tagging": <0-100 or null>,
    "track_photon_selection": <0-100 or null>,
    "pid": <0-100 or null>,
    "fits_and_constraints": <0-100 or null>,
    "mc_and_decay_card": <0-100 or null>
  },
  "fatal_mismatch": true | false,
  "verdict": "<one sentence>"
}
\end{promptbox}

\section{BESIII Benchmark Data Description}
\label{app:benchmark}
\subsection{Construction Protocol for BESIII-PhysicsQA}

BESIII-PhysicsQA comprises 1,000 retrieval QA pairs, covering 756 published BESIII papers. Construction follows a deterministic seven-step protocol. Its core principle is that the language model determines no facts: hop chains, gold facts, and answer values are all produced by deterministic queries over a prebuilt fact index. The model participates only in structural annotation and question rewriting, and outputs only line numbers or question strings.

\textbf{Structural annotation.} Each paper is annotated with section spans, canonical types, independent discussion units, and table attribution. The model outputs only line numbers, making text-block identifiers a pure function of the frozen artifact and gold locations invariant across reconstructions. The whole corpus is annotated uniformly to avoid different length distributions between gold and distractor blocks.

\textbf{Text-block materialization and fact extraction.} Text blocks are split according to annotations, length budgets are applied by canonical section type, and each table is promoted to a separate text block. Facts are also extracted, and an index and construction report are generated. No model is involved in this step.

\textbf{Deterministic item selection and assembly.} Realistic scientific task scenarios are written first; items are discarded if none can be written. Candidate anchors are queried by category and passed through template-based gates and a Monte Carlo generator consensus gate. Answers are computed by query; systematic-error cases are arithmetic cross-checked, not model-generated.

\textbf{Isolation validation.} Hop count is defined as the minimum text-block set cover over all gold facts, solved exactly. Items are rejected one by one for cases such as a single block containing all gold facts, answer-value leakage, excessively low anchor selectivity, and arithmetic non-reproducibility. Physics papers are highly self-contained; within-paper multi-hop structure is far less common than it appears, so isolation is verified item by item against real text blocks.

\textbf{Natural-language rewriting.} This is the only step that calls the model per item. Inputs are the scenario, fixed hop chain, gold facts, answer, and eight hard checks; only a question string is requested. Checks include:
\begin{itemize}
  \item prohibit reverse lookup of papers from known parameters;
  \item restrict length and a single main condition;
  \item require a scientific-task motivation;
  \item prohibit artificial hop-chain elongation;
  \item prohibit leaking locations or arithmetic;
  \item prohibit pipeline meta-vocabulary;
  \item prohibit rare-parameter conjunctions as identifiers;
  \item require a unique, machine-verifiable answer.
\end{itemize}
If any check fails, rewrite once; if it fails again, discard the item.

\textbf{Adversarial validation.} Gold text blocks are hidden and only retrieved distractors are provided; if the model still answers correctly, the item does not test retrieval and is discarded.

\textbf{Export.} At export, clustering keys are enforced and the correspondence between QA pairs and the index is frozen.

\end{document}